# Accuracy vs. Computational Cost Tradeoff in Distributed Computer System Simulation

Adrian Colaso
Computer Engineering Group
University of Cantabria
Spain
colasoa@unican.es

Pablo Prieto
Computer Engineering Group
University of Cantabria
Spain
prietop@unican.es

Jose Angel Herrero
Computer Engineering Group
University of Cantabria
Spain
herreroja@unican.es

Pablo Abad
Computer Engineering Group
University of Cantabria
Spain
abadp@unican.es

Valentin Puente
Computer Engineering Group
University of Cantabria
Spain
vpuente@unican.es

Jose Angel Gregorio
Computer Engineering Group
University of Cantabria
Spain
monaster@unican.es

## ABSTRACT

**Simulation is a fundamental research tool in the computer architecture field. These kinds of tools enable the exploration and evaluation of architectural proposals capturing the most relevant aspects of the highly complex systems under study. Many state-of-the-art simulation tools focus on single-system scenarios, but the scalability required by trending applications has shifted towards distributed computing systems integrated via complex software stacks. Web services with client-server architectures or distributed storage and processing of scale-out data analytics (Big Data) are among the main exponents. The complete simulation of a distributed computer system is the appropriate methodology to conduct accurate evaluations. Unfortunately, this methodology could have a significant impact on the already large computational effort derived from detailed simulation. In this work, we conduct a set of experiments to evaluate this accuracy/cost tradeoff. We measure the error made if client-server applications are evaluated in a single-node environment, as well as the overhead induced by the methodology and simulation tool employed for multi-node simulations. We quantify this error for different micro-architecture components, such as last-level cache and instruction/data TLB. Our findings show that accuracy loss can lead to completely wrong conclusions about the effects of proposed hardware optimizations. Fortunately, our results also demonstrate that the computational overhead of a multi-node simulation framework is affordable, suggesting multi-node simulation as the most appropriate methodology.**

## 1 INTRODUCTION

The evolution of computing systems into the extremely complex structures available nowadays has obliged the development of increasingly sophisticated simulation frameworks. Simulation tools have in many cases become complex pieces of software exceeding 100,000 code lines and mixing multiple programming languages [1]. Under these circumstances, the tradeoff between accuracy, computational effort and tool complexity is hard to maintain. Working with highly detailed tools permits in most cases the evaluation of the whole software stack (i.e., Operating System, Runtime and Application) and exposes low-level micro-architecture details for evaluation. Unfortunately, it frequently requires a steep learning curve [2] and/or could lead to unaffordable simulation time [3].

Additionally, conventional and widely employed benchmark suites such as SPEC CPU2006 [4] and PARSEC [5] are progressively becoming less representative of many real usage scenarios [6]. In contrast, applications for distributed data storing, processing and serving or web services responding to client-server architecture are becoming dominant in the software market. The cost constraints, complexity and scalability of this kind of applications have obliged the abstraction level to be raised by the programmer. Therefore, a highly complex software stack, in many cases in the form of distributed frameworks, is mandatory.

Simulating multiple nodes increases the complexity of many aspects concerning the evaluation process. Additional simulated components mean longer execution times and more required resources to perform simulation. For these reasons, the cost/accuracy tradeoff becomes even more fragile. Despite the inherent distributed nature of such benchmarks, we could be tempted to make use of single-node system simulation as evaluation methodology (all benchmark elements running in the same simulated node). In this case, we will be aggregating a new source of error, caused by the interference of all benchmark elements sharing the same hardware. The risk of such approaches is that they introduce a non-negligible deviation from real system behavior that could lead to large error margins [7][8][9] affecting the conclusions derived from their use.

The main target of this work focuses on the quantification of the error caused by the interaction of the different software pieces running on top of the same hardware. This way we will estimate whether the increased complexity of simulating additional nodes is avoidable or, in contrast, mandatory. To do so, we have conducted a group of experiments comparing single-node and multi-node execution of distributed applications on both real and simulated environments. Additionally, we carry out an analysis of the simulation overheads caused by the increased complexity when the appropriate methodology is employed.

We present our findings and experiment results with the following structure: Section 2 revises some research works concerning distributed systems evaluation and related to the content of this paper. Section 3 performs a detailed description of the evaluation framework employed, including benchmarks, simulation tools, workload configuration, etc. Section 4 describes the experiments and results performed to quantify the error of inappropriate methodologies. Section 5 evaluates the overheads of multi-node simulation, providing deeper insight on the accuracy/cost tradeoff between single-node and multi-node simulation. Finally, Section 6 states our main conclusions and describes the future work lines.

## 2 DISTRIBUTED COMPUTING EVALUATION

The relevance of non-conventional benchmarks targeting client-server applications with multiple execution nodes is visible in the increasing number of studies devoted to the analysis of this kind of applications. The literature provides benchmarks covering a wide range of multi-node scenarios. Some benchmarking suites target specific software, such as Hadoop [10] environments (Hibench [11]) or NoSQL databases (YCSB [12]). In contrast, alternative benchmark suites such as CloudSuite [6] and BigDataBench [13] cover a much wider range of domains, such as offline analytics, real-time analytics and online services. Until now, many of the research works making use of these benchmarks have focused on the microarchitectural characterization of these applications [14][15][16][17][18][19], relying on methodologies based on performance profiling tools such as perf [20] or VTune [21].

The scope of profiling tools is mainly limited to characterization, while testing the functionality and performance of new hardware mechanisms requires the utilization of simulation tools such as the ones employed in this work. Some recent works have already explored the utilization of gem5 for the evaluation of client-server applications. The authors in [22] describe how to run client-server benchmarks in x86 dual-system mode on gem5. To do so, they perform the required modifications to gem5 original code in order to set up a dual system communicating through an Ethernet link. In [23], the authors have developed a fully-distributed version of gem5 (dist-gem5), supporting the simulation of multiple nodes on multiple physical hosts (one host for each simulated node). Communication among nodes is performed through the real network, usually Ethernet. The speedup and scalability of dist-gem5 is evaluated by simulating up to 63-node cluster.

In this work we have combined the utilization of emerging benchmarking suites such as SPECweb [24] and the Yahoo! Cloud Serving Benchmark(YCSB) [12] with a gem5 implementation with a similar functionality to dist-gem5. In contrast to the dist-gem5 approach, our gem5 framework also simulates communication elements, running every simulated piece on a single physical node. This is a similar approach to the one employed in [22]. To the best of our knowledge, this is one of the few works where multi-node applications are evaluated making use of a full-system simulation running the whole software stack.

## 3 EVALUATION FRAMEWORK

In this section, we present a detailed description of the tested software and simulation tools, as well as our experimental methodology.

### 3.1 Distributed Applications

In order to work with applications more representative of a real scenario, we will make use of the SPECweb [24] and the Yahoo! Cloud Serving (YCSB) [12] benchmarks for all the experiments in this document. Making use of these two benchmarks we are able to generate 15 different workloads for evaluation, summarized in Table 1 and briefly described in this section.

SPECweb is a complex benchmark developed to assist in the performance evaluation of web servers. It has four major components (prime-client, clients, web-server and back-end simulator) that correspond to a client-server architecture design, suitable for multi-node environments. The benchmark clients generate HTTP requests to the server and receive responses. Their behavior is controlled by the prime client. The Web server is the collection of hardware and software in charge of managing client requests. Finally, the back-end simulator emulates an application server that the Web server must communicate with in order to complete HTTP responses. Three different workloads are provided, matching different realistic use scenarios: banking, e-commerce and support. Banking workload, related to online banking, is based on the study of web server logs from a major Texas bank. The E-commerce workload simulates a Web server that sells computer systems. Finally, Support workload simulates a vendor´s support web site.

**Table 1. Workloads**

| Bench | Workload | Description |
|---|---|---|
| SPECweb | BANK | Online banking. Rampup=60 seconds, Warmup=60 secs, Run = 300 secs. |
| | E-COMM | Sales web. Rampup=60 seconds, Warmup=60 secs, Run= 300 secs. |
| | SUPP | Vendor's support. Rampup= 60 secs, Warmup=60 secs, Run=300 secs. |
| YCSB + (Cassandra/MongoDB) | WA | Update heavy. 50% reads, 50% writes. Record count = 1000 |
| | WB | Read mostly. 95% reads, 5% writes. Record count = 1000 |
| | WC | Read only. 100% read. Record count = 1000 |
| | WD | Read latest. 95% read, 5% insert. Record count = 1000 |
| | WE | Short ranges. 95% scan, 5% insert. Record count = 1000 |
| | WF | Read-Modify-write. 50% read, 50% r-m-w. Record count = 1000 |

YCSB is a benchmarking framework to assist in the evaluation of cloud systems. It consists of a workload-generation client (with six different standard workloads) and a Database interface layer to connect to different cloud serving stores. The core package includes a set of pre-defined workloads that try to model different applications, such as picture tagging, user status updates or threaded conversations [12]. These workloads are described in Table 1. The database tested, generated with YCSB, consists of several million records, for a total size of over 12GB.

The first serving database is Apache Cassandra [25], a popular Java implementation of a column-family NoSQL database. This system has been designed to work with large data volumes on top of commodity hardware, providing high availability and fault tolerance features. Cassandra makes use of its own query language (Cassandra Query Language or CQL) and also provides Hadoop integration. Nowadays, more than 600 companies employ Cassandra software, according to [26]. The second data-management application, MongoDB [27], is a document-oriented database designed to provide scalability, performance and high availability. Documents are stored as binary JSON objects, supporting field and range queries as well as regular expression searches. Data distribution across multiple machines is implemented through sharding, while high availability and fault tolerance are implemented through replica sets. As in the case of Cassandra, MongoDB is also one of the most popular document stores, with a great diversity of users [26].

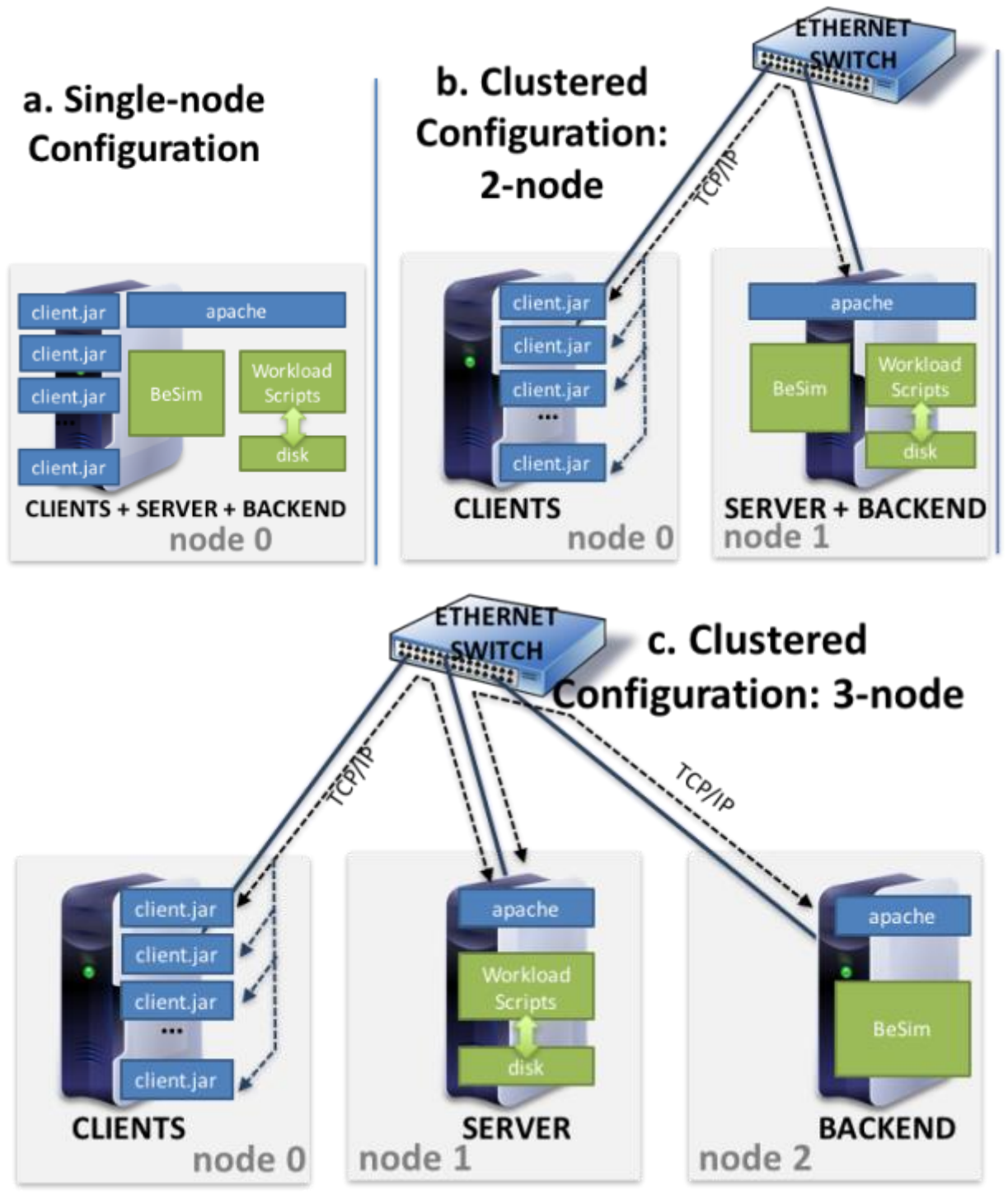

**Figure 1. Multi-node structures simulated with gem5**

## 3.2 Simulation Framework

The structure of the proposed experiments imposes tough requirements on the simulation framework, such as detailed full-system hardware description and multi-node simulation. Gem5 [1] implements the necessary features and therefore, is the tool that we choose to conduct our evaluations. This enables the evaluation of clustered configurations such as the ones shown in Figure 1 and employed for SPECweb simulation (Section 6).

To achieve multi-node simulation at an affordable cost (simulation time) only accurate hardware simulation is used during the execution of a significant fraction of the Region Of Interest (ROI). In the case of the applications under evaluation, the records needed to generate both Cassandra and MongoDB databases as well as the warm-up phase of SPECweb should be discarded, but still requires several minutes to complete in a real system. To reduce the simulation time, we follow a two-step approach to generate our workloads, as described in Figure 2. First, we “fast-forward” applications to the region of interest, taking checkpoints once the application reaches this state (a checkpoint includes the architectural state of processor and memory and it is kept in persistent storage). Afterwards, these checkpoints are employed as workloads, loaded with detailed architectural simulation.

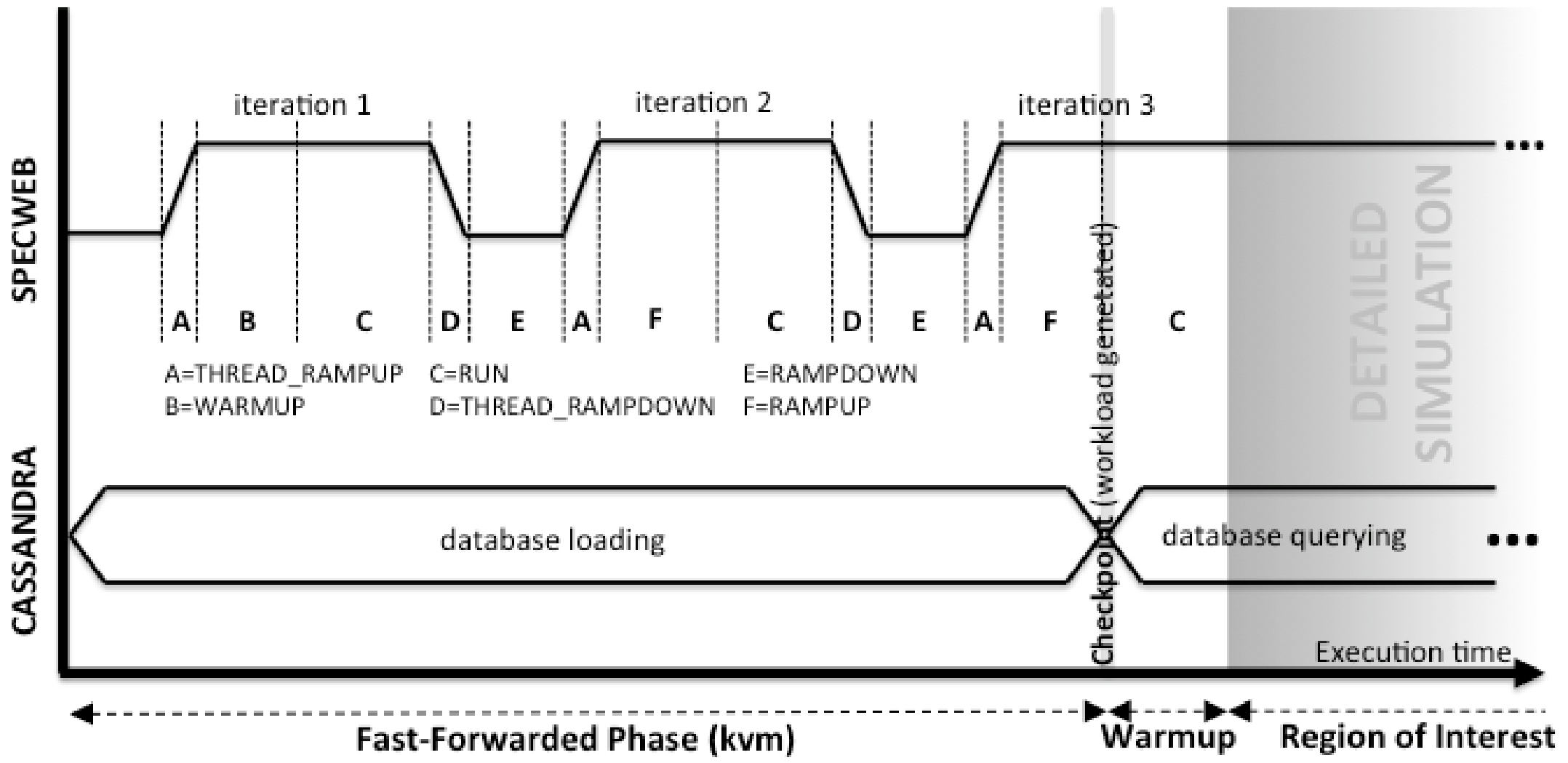

**Figure 2. Workload generation process for SpecWeb and YCSB. Each application is fast-forwarded to the ROI making use of kvm-assisted execution.**

### 3.3 Hardware & Software Stack Configuration

We run all our experiments (single-node and distributed configurations) on both real hardware and full-system simulation framework. Physical nodes have two Intel Xeon X5650 chips running at 2.67 GHz (24 threads) and a main memory of 48 GB. In the multi-node setup, nodes are connected through a 1Gbps Ethernet network. We access the Performance Monitoring Unit (PMU) of the processor through the Linux perf tool [20]. Simulation-based experiments mimic the micro-architecture configuration of the real machines, only scaling down the number of processor cores from 12 to 4 to speedup simulations.

The complete software stack is employed for evaluation in both real and simulated environments. SPECweb and YCSB benchmarks run on top of a Linux OS (Debian 8 distribution, kernel version 3.18.34) + Oracle’s open source Java Virtual Machine, 1.5 and 1.7 versions respectively.

Multiple runs are always used to ensure that we fulfill a strict 95% confidence interval. Although some memory models (ruby) support trace driven warm-up from the checkpoints, we used the gem5 classic memory model, which does not do so. Starting from a checkpoint, the memory hierarchy is warmed up during enough cycles before starting to collect statistics.

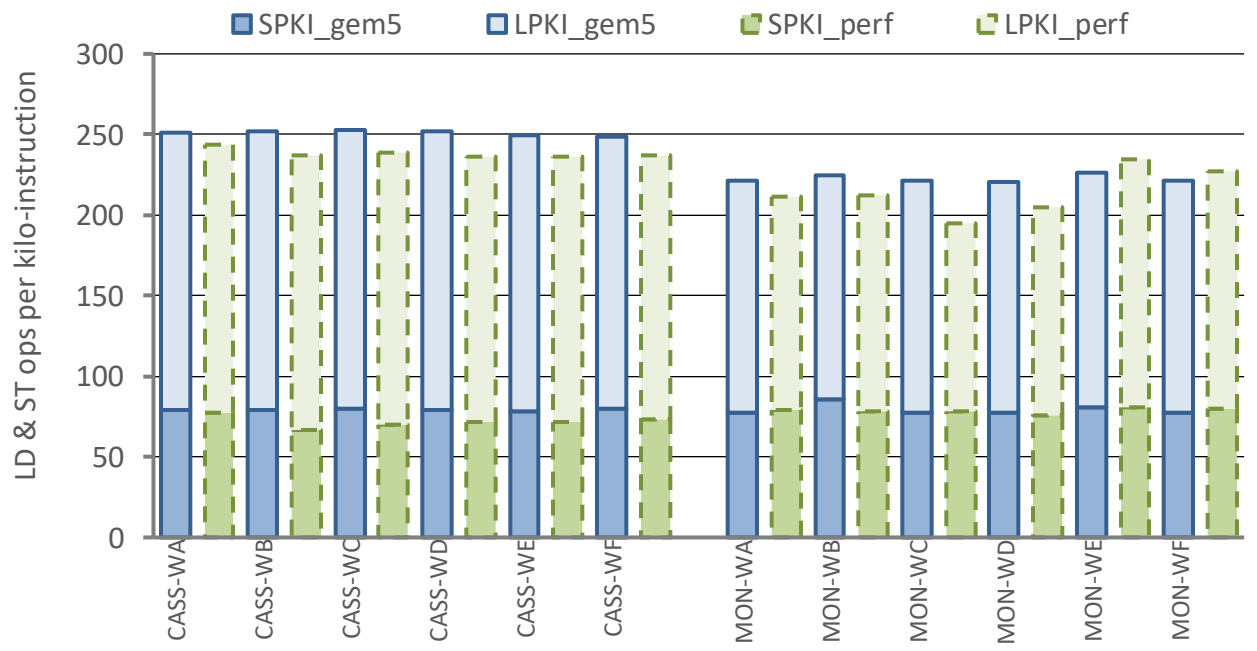

**Figure 3. Load and Store profile (events per kilo-instruction) making use of two different execution environments (real hardware vs. gem5)**

### 3.4 Simulation Framework Validation

As our experiments combine the utilization of PMU profiling and Full-system simulation, we carry out a preliminary experiment trying to validate the joint utilization of both methodologies. To do so, we will make use of YCSB workloads running on both environments and measuring the load/store footprint of each environment. In both cases, real hardware and gem5, the database content is replicated generating a 1GB database with the same YCSB commands. We run a fixed number of records (one thousand) on both Cassandra and MongoDB databases and measure the fraction of Load and Store operations for each workload. Figure 3 shows the results obtained. The y-axis represents the fraction of Load and Store operations for each one thousand instructions executed. Solid blue bars represent simulated results (gem5) while dotted green bars show the numbers obtained with perf. As can be seen, deviation is minimal, less than 5% on average.

# 4 EXPERIMENTAL RESULTS

In this section we conduct multiple experiments evaluating different components of the memory hierarchy, as well as overall system performance. We measure the miss rate of L1 instruction cache, data and instruction translation lookaside buffer and last level cache under two different scenarios, client+server running on the same node against client and server running on two different nodes.

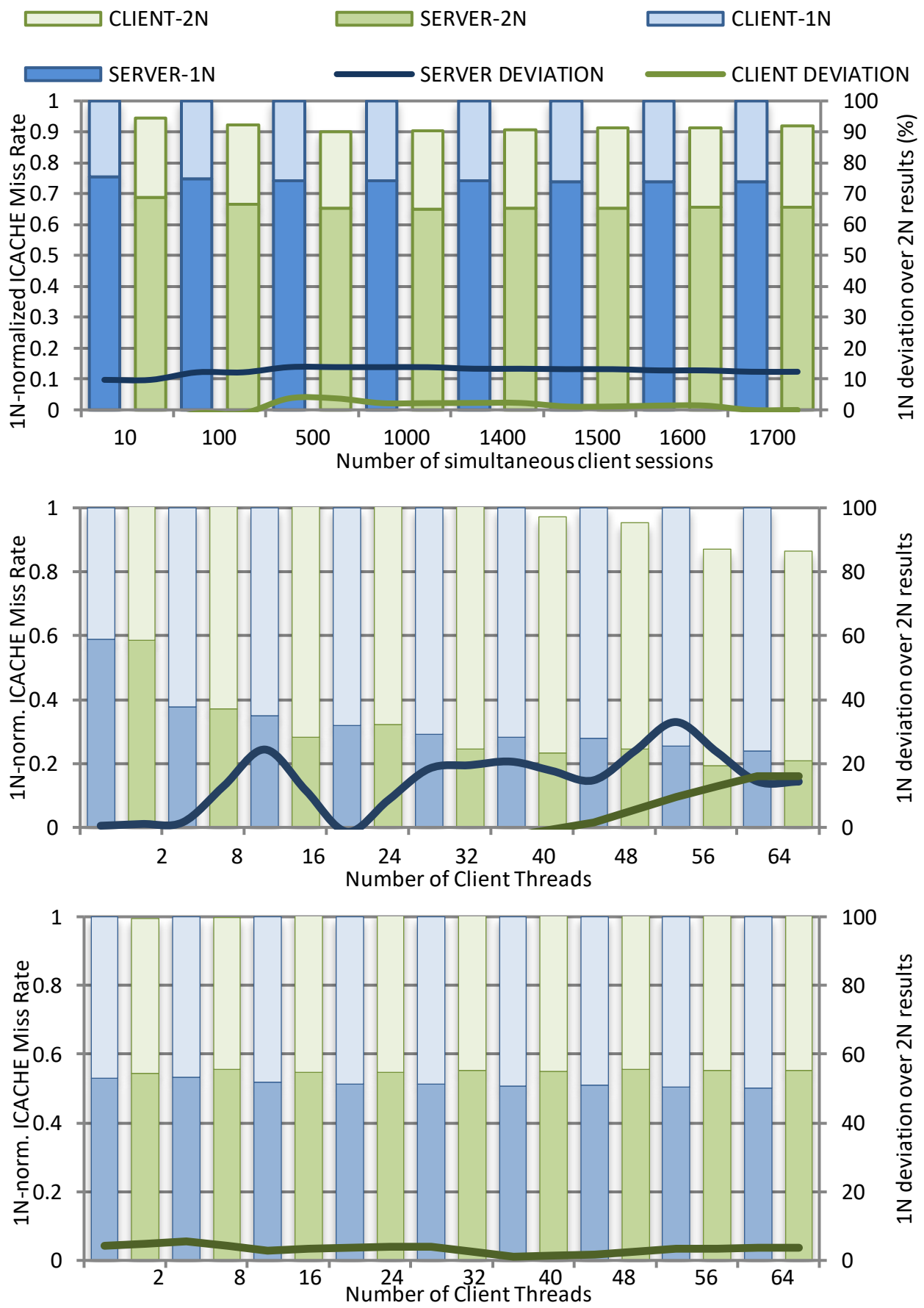

**Figure 4. Evolution of ICACHE miss rate (normalized to 1-node values) as the number of simultaneous client sessions grows. (up) SPECweb support results. (mid) Cassandra, WA results. (down) MongoDB, WA results. Miss rate values (bars) and individual client/server deviation (lines)**

## 4.1 L1I Cache and Translation Lookaside Buffer

Figure 4 and Figure 5 present L1I cache and TLB miss-rate results for client and server running on the same node (1N) and independent client and server execution (2N). The evolution of Client and Server Miss Rate (normalized to 1-node results), as the number of simultaneous client sessions/threads ramps up, is provided. For the sake of simplicity only results corresponding to one workload of each benchmark suite are provided. This way, Figure 4.up shows results for the support workload from the SpecWeb benchmark, while Figure 4.mid and Figure 4.down corresponds to Workload A from YCSB (Cassandra and MongoDB respectively). The values of this section have been obtained measuring PMU features on real hardware.

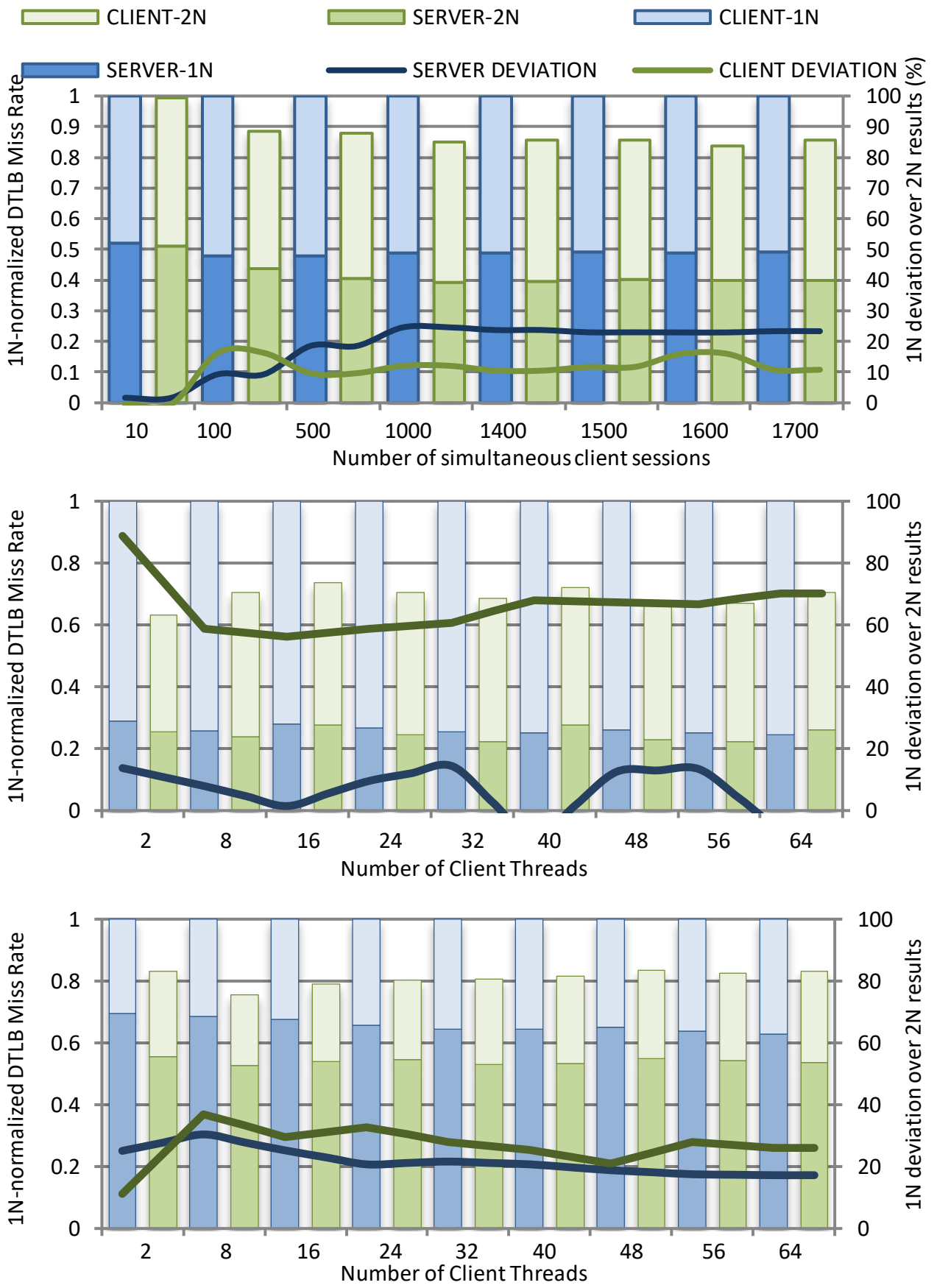

**Figure 5. Evolution of DTLB miss rate (normalized to 1-node values) as the number of simultaneous client sessions grows. (up) SPECweb support results. (mid) Cassandra, WA results. (down) MongoDB, WA results. Miss rate values (bars) and individual client/server deviation (lines)**

ICACHE results show that only a marginal deviation is caused by the joint execution of client and server code. On the client side, deviation is nearly imperceptible. In contrast, on the server side this difference grows up to 20% in the most adverse cases, such as YCSB querying Cassandra database. The main reason behind this difference could be in the dissimilar size of client and server instruction working sets. While clients of these benchmarks are "simple" synthetic request generators, serving applications are known by the large working set associated to their instructions.

For those results concerning TLB in Figure 5, we observe how the error caused by single-node execution increases. In this experiment deviation moves to values ranging from 20% to 70%. Main deviation is observed on the client side, but also server results consistently exceed 20%. These values are far from negligible, suggesting that single-node simplification might lead to incorrect conclusions. To confirm these observations we extend our experiments to alternative elements of the memory hierarchy, such as the Last Level Cache.

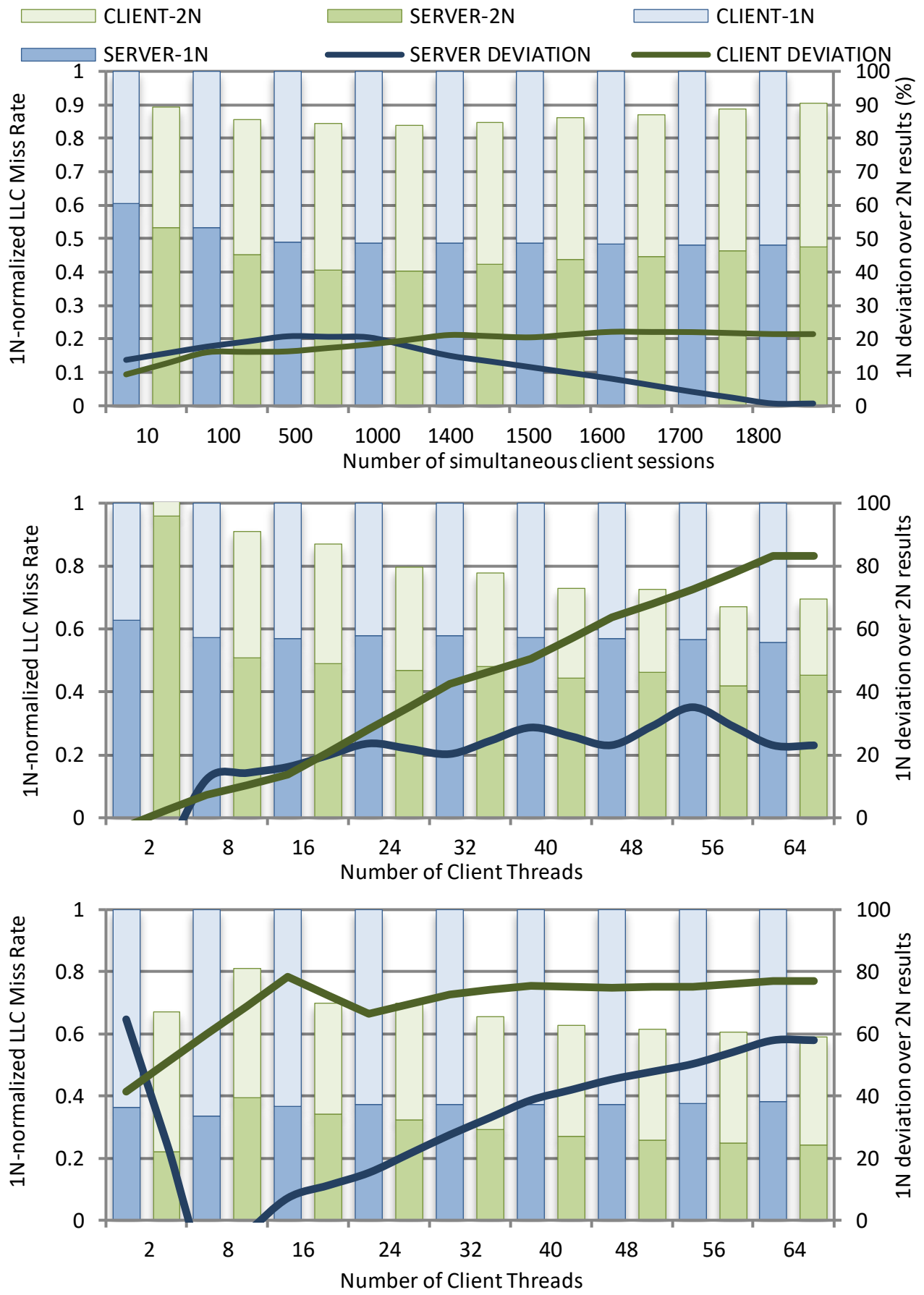

**Figure 6. Evolution of LLC miss rate (normalized to 1-node values) as the number of simultaneous client sessions grows. (up) SPECweb support. (mid) Cassandra WA. (down) MongoDB WA. Miss rate values (bars) and individual client/server deviation (lines)**

## 4.2 Last Level Cache

We extend our evaluation to LLC performance under single-node and multi-node configurations, because this is one of the components where client-server interaction becomes more evident. For the first LLC experiment we employ the same methodology (PMU features) as in previous section, presenting Figure 6 LLC miss-rate results as the number of clients grows for client and server running on the same node (1N) and independent client and server execution (2N). As can be seen, the deviation for 1N results from the "realistic" (2N) scenario is much more significant than the observed results in L1I, ranging from 20% (SPECweb) to 80% (Cassandra and MongoDB) for server LLC accesses and from 20% (SPECweb and Cassandra) to 60% (MongoDB) in the case of benchmark clients.

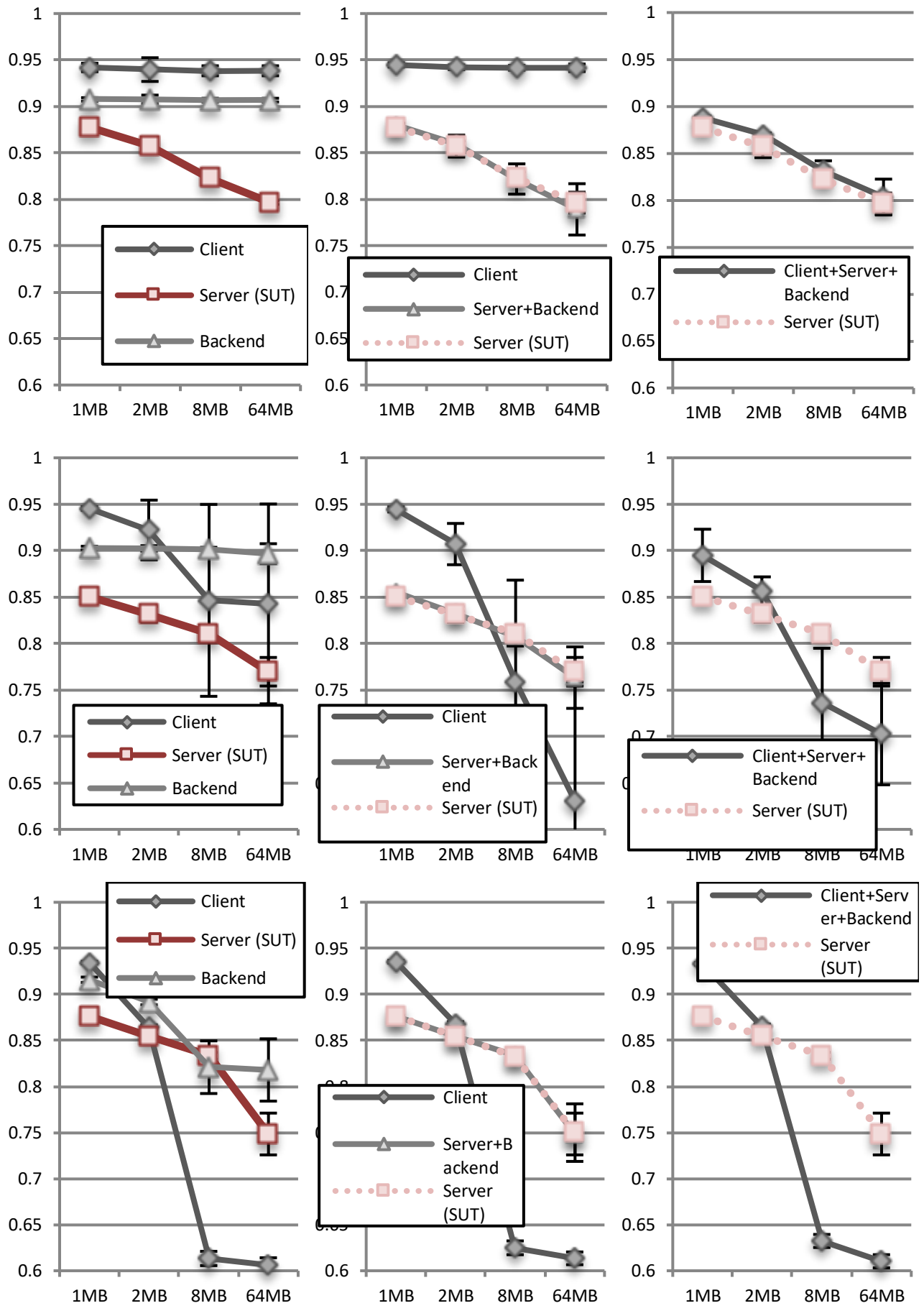

**Figure 7. Miss rate evolution with LLC size. (left column) 3-node simulation, (mid column) 2-node simulation and (right column) 1-node simulation. Banking (up row), Ecommerce (mid row) and Support (down row).**

Given the adverse effect that these deviation values could have on performance estimations, we decided to extend the LLC evaluation. Allowed by the proposed simulation framework and methodology, we focus our attention on the performance impact of variable LLC size. Making use of the simulated environment we are able to explore how the deviation evolves as the LLC size grows from 1MB to 64MB. It must be noticed that this kind of experiment is only available through simulation frameworks like the one employed here or those proposed in [22][23].

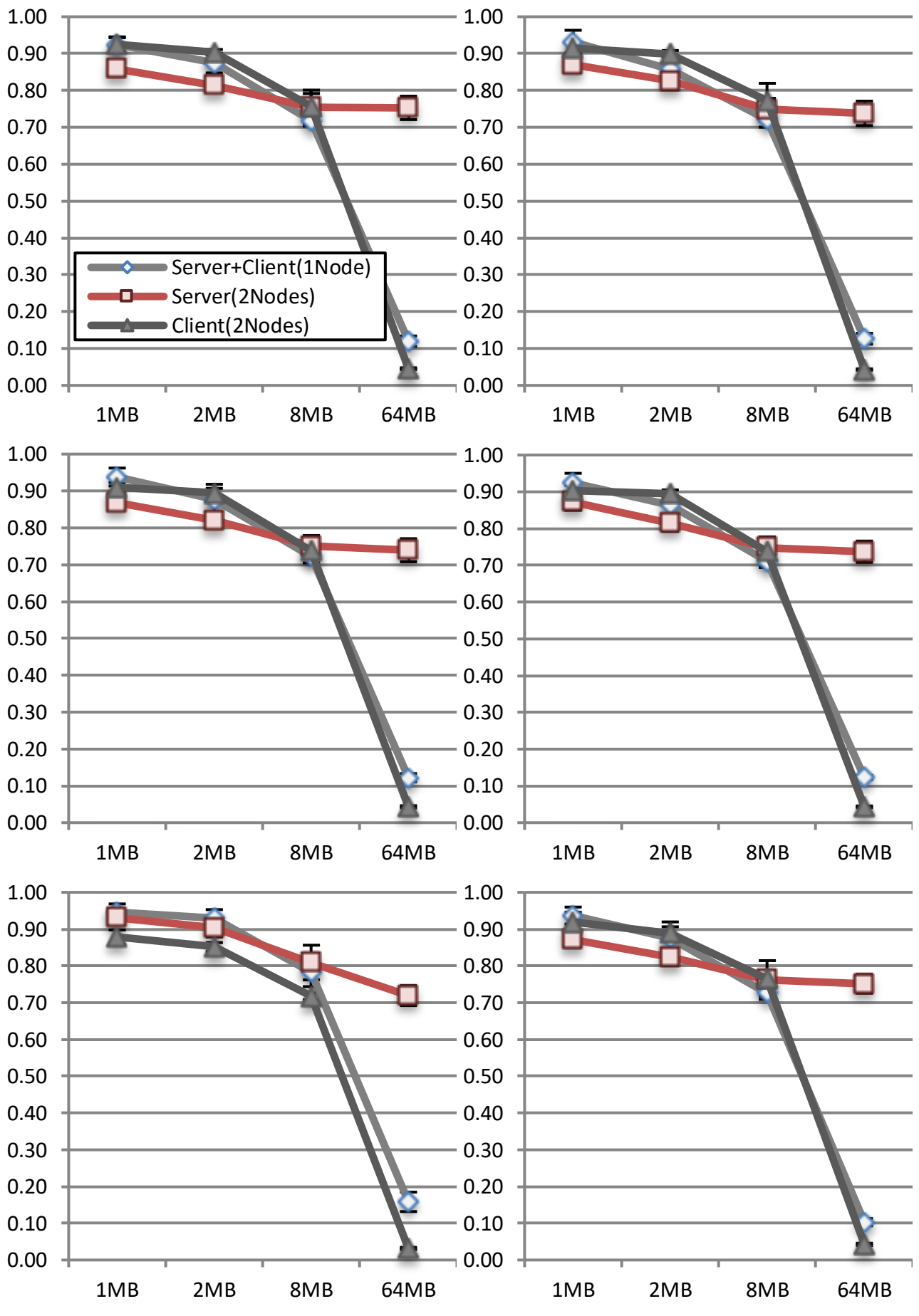

**Figure 8. Miss rate evolution with LLC size (YCSB + Cassandra). 1-node vs. 2-node results, workloads A to F.**

According to the SPECweb deployment guide, the System Under Test (SUT) corresponds exclusively to the Web Server (node 1 in Figure 1.c). We analyze the effect of fusing the remaining components and their interference on LLC performance. Figure 7 shows miss-rate evolution in LLC as the size of this component is increased. Each row in the figure corresponds to the three workloads from Spec-web benchmark, Banking (up) E-commerce (mid) and Support (down). The left column in Figure 7 provides SUT results simulating a 3-node distributed system, as sketched in Figure 1.c. This could be considered as the closest configuration to a real scenario. Central and right columns show the results obtained for the same metric (LLC miss-rate) as configuration is gradually simplified to simulate just a single-node. Dotted lines in central and right columns represent the results obtained for SUT on a realistic 3-node system. The deviation of solid lines from this reference dotted line shows the error caused by inadequate simulation. It can be observed that this error strongly depends on the workload, ranging from less than 5% (Banking) to more than 20% (Support). This deviation could be defined as the error made if we try to evaluate Server behavior with a single-node system (all software pieces running on the same node).

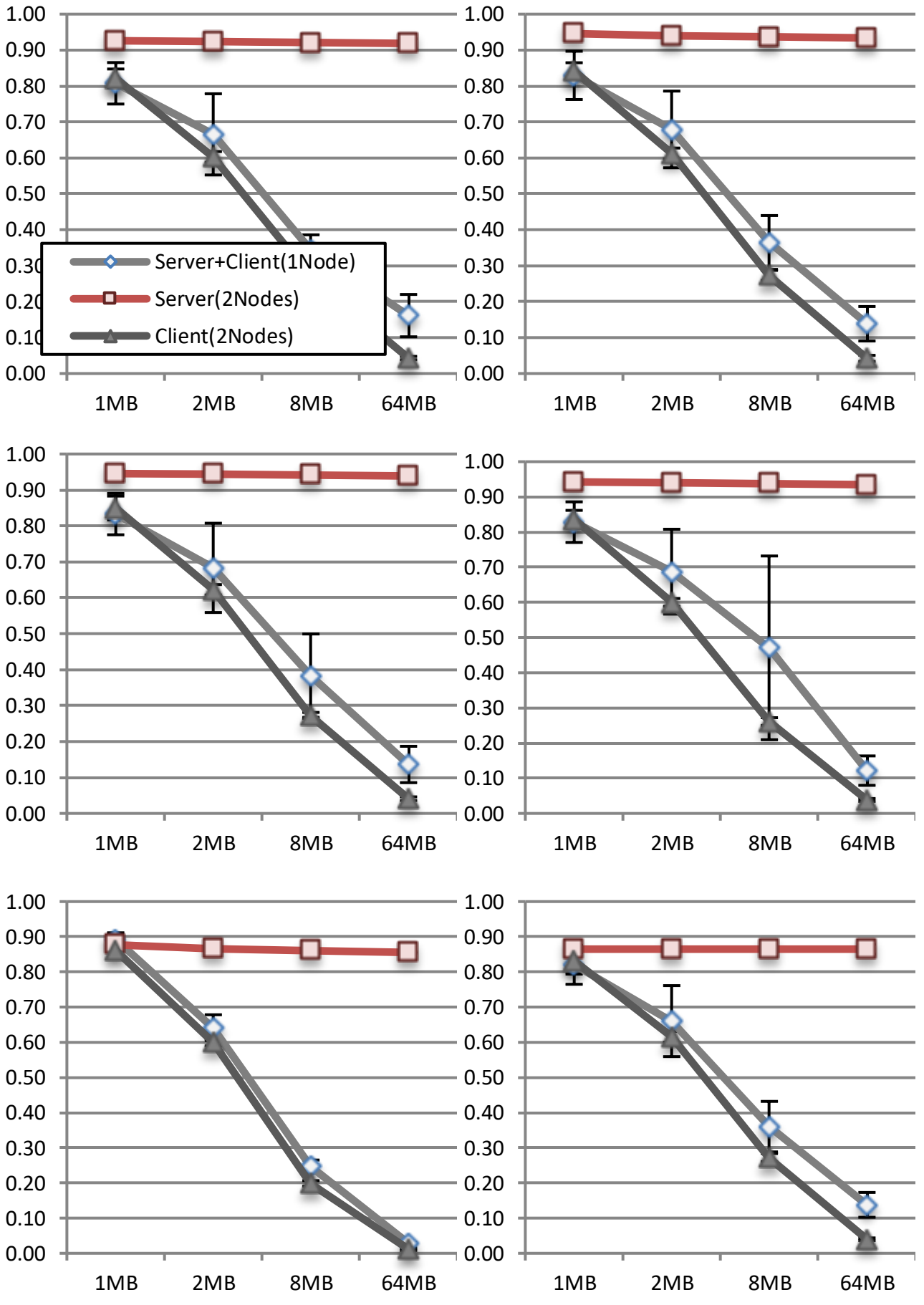

**Figure 9. Miss rate evolution with LLC size (YCSB + MongoDB). 1-node vs. 2-node results, workloads A to F.**

Figure 8 and Figure 9 present a similar study for Cassandra and MongoDB workloads respectively. In this case, we limit our evaluation to 1-node and 2-node configurations. For all the workloads analyzed results show that miss-rate curve is shaped by the memory operations performed by the client. Server+Client (1-Node system) and Client (2-Node system) results follow a similar progression, miss-rate reducing as LLC size increases. In contrast, if we analyze the standalone behavior of server side, we observe that ***cache size is not able to improve miss-rate values*** for the range of LLC capacities evaluated. Making use of the correct methodology (2-Node system, server running standalone), our results show that LLC size has few or none influence on miss rate. In contrast, miss rate in single-node systems shows a great dependency on LLC size, which is completely false because server behavior has been completely masked by client cache accesses. In this case, The main reason for these results is the unbalanced working-set size of the code running on Client and Server sides. Client code, corresponding to the YCSB benchmark is merely in charge of generating the requests to the database, which requires only a moderate working-set. Most of the LLC accesses corresponding to the client end-up with a hit. On the other hand, the Server side runs the software implementing the database engine, which works with randomized queries to a 1GB data set. Such a large working-set obtains few benefit from a LLC increase from 1MB to 64MB.

This last group of results is useful to state the main contribution of this work. The potential error caused by a simplified simulation methodology is not irrelevant and has led to incorrect conclusions. In this particular case, the decision of provisioning the system with cores with larger cache, based on the single-node evaluation, would have been completely useless. If we try to evaluate a micro-architectural proposal to improve the performance of a database engine, single node simulation is an unsuitable methodology. Running server and client code on the same hardware leads to a load/store behavior mostly dominated by the client side. This means that it is difficult to evaluate the effects of any proposal on server performance, it being necessary to adopt more elaborated simulation methodologies such as the one employed in this paper or those proposed in [23][22].

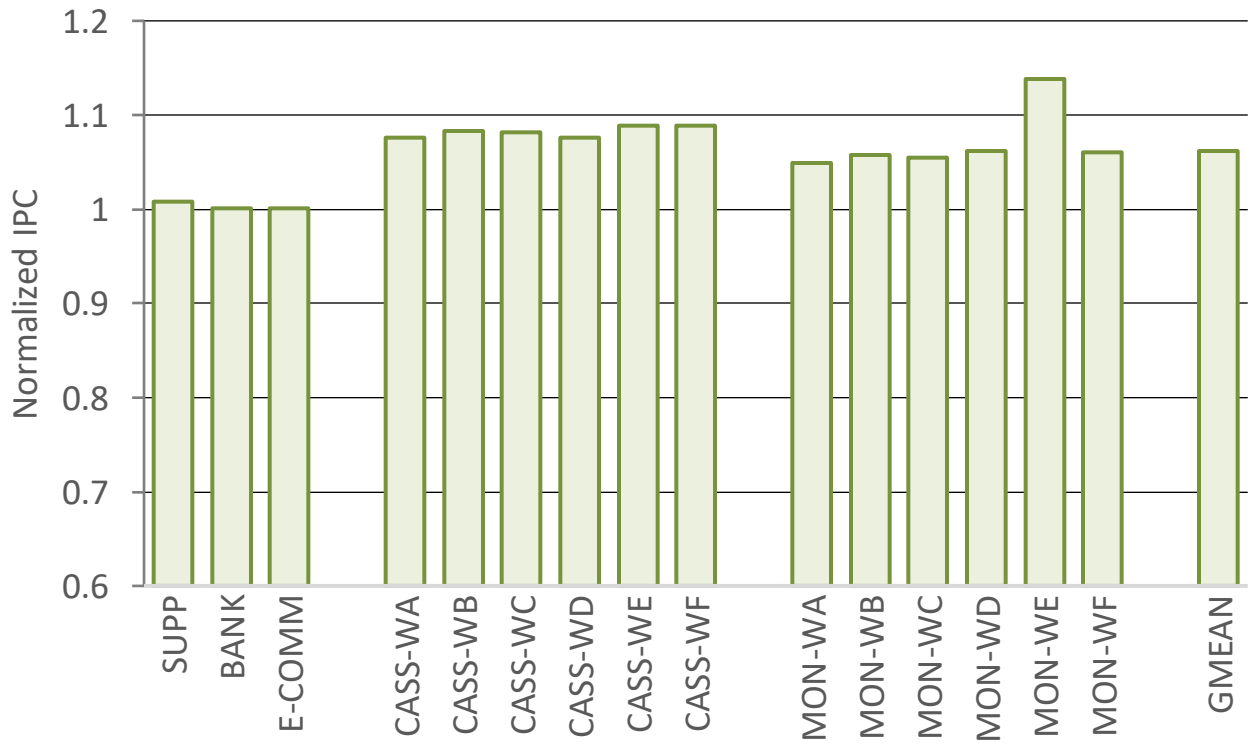

**Figure 10. Client + Server (1-node simulation) IPC normalized to server standalone results.**

### 4.3 Overall System Performance

We complete this section measuring the overall IPC deviation caused by client-server interaction in single-node system. The same system configuration is maintained, setting LLC size to 8MB in this experiment. Figure 10 represents the IPC values of single-node system normalized to the IPC obtained by the server running standalone. As can be seen, client-server interaction modifies IPC results, reaching deviation values up to 15%. IPC deviation seems to be much more relevant in the case of YCSB workloads, caused by a more dissimilar behavior between client and server performance (total IPC is artificially increased by the higher IPC values observed in the client). We can conclude that a 10% error would be added to performance evaluation process merely by letting client and server code share LLC resources, which is not representative of a real scenario.

## 5 SIMULATION OVERHEAD

This last section analyzes the computational overhead added by distributed system simulations. We evaluate the computational effort required for correct client-server application benchmarking, as well as the execution time overhead and the additional memory footprint of dual-node simulation compared to a single-node configuration.

Even with the fastest non-assisted simulation mode in gem5 (i.e. atomic), database generation and SPECweb warm-up would require an unfeasible amount of time. Therefore, multi-threaded virtual-machine assisted fast-forward is a key element in our tool. Given the coarse synchronization between simulator and virtual machine monitor, and the use of gem5 support for multithread event queues, this setup allows the application to be “fast-forwarded” to the interest point at near-native speed (if the running server has the same number of cores than the simulated machine).

We have conducted an experiment to measure the time overhead required to reach the Region Of Interest for the YCSB-related workloads (for checkpoint generation) under three execution modes: real, kvm and atomic. This process requires to complete the required number of database queries to fully load a Cassandra database. We provide in Figure 11 the results obtained for two different database sizes, 10MB and 1GB. The x-axis represents the required completion time in minutes (10MB) or hours (1GB). As can be seen, both kvm and atomic present a large execution-time overhead with respect to real hardware. Making use of atomic execution mode, even the generation of a non-realistic database (10MB) requires nearly a week to be completed. In contrast, this time is reduced to barely 10 minutes through hardware-assisted simulation (kvm-multithread). The big difference is that making use of hardware-assisted simulation is the only way to allow for the generation of a 1GB database at an affordable cost, less than 24 hours. Once the database is completely loaded a checkpoint can be taken, so this process only needs to be done once. If we try to generate the same database through atomic simulation mode, the execution time will increase up to a year, which is a completely unrealistic delay (it should be noted that the atomic results for the 1GB database have been extrapolated from 10MB results, given the inability to perform such long simulations). Similar results have been obtained for the SPECweb warm-up phase, as hardware-assisted simulation is mandatory in both cases.

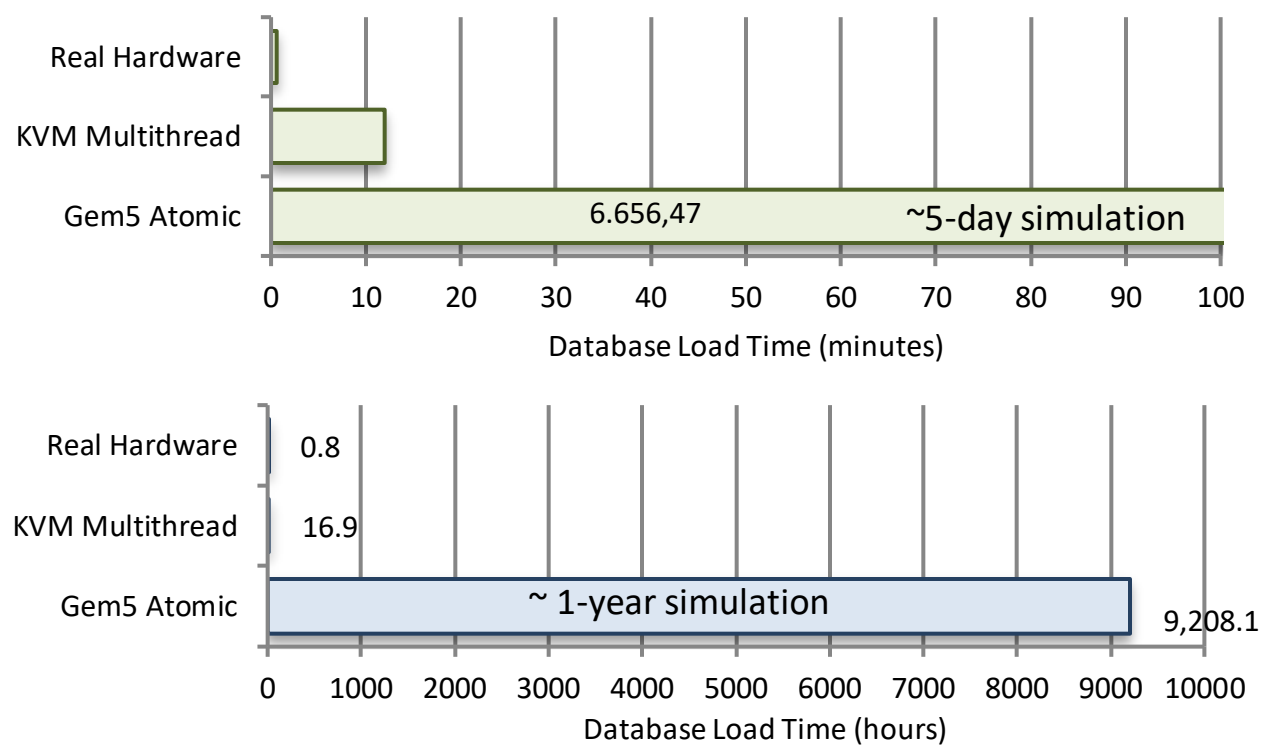

**Figure 11. Database load time (checkpoint generation) with different methodologies, 10Mbyte (up) and 1Gbyte (down).**

The results in Figure 12 have been obtained from the simulations performed in previous sections. Figure 12.up represents the ~~additional~~ simulation time required to complete the same task in a dual-node system when compared to an equivalent simulation of a single-node system. In both cases the number of total cores simulated is the same. As can be seen, the addition of a second node does not incur a simulation time increase. The effect of the extra simulated node over execution time is negligible.

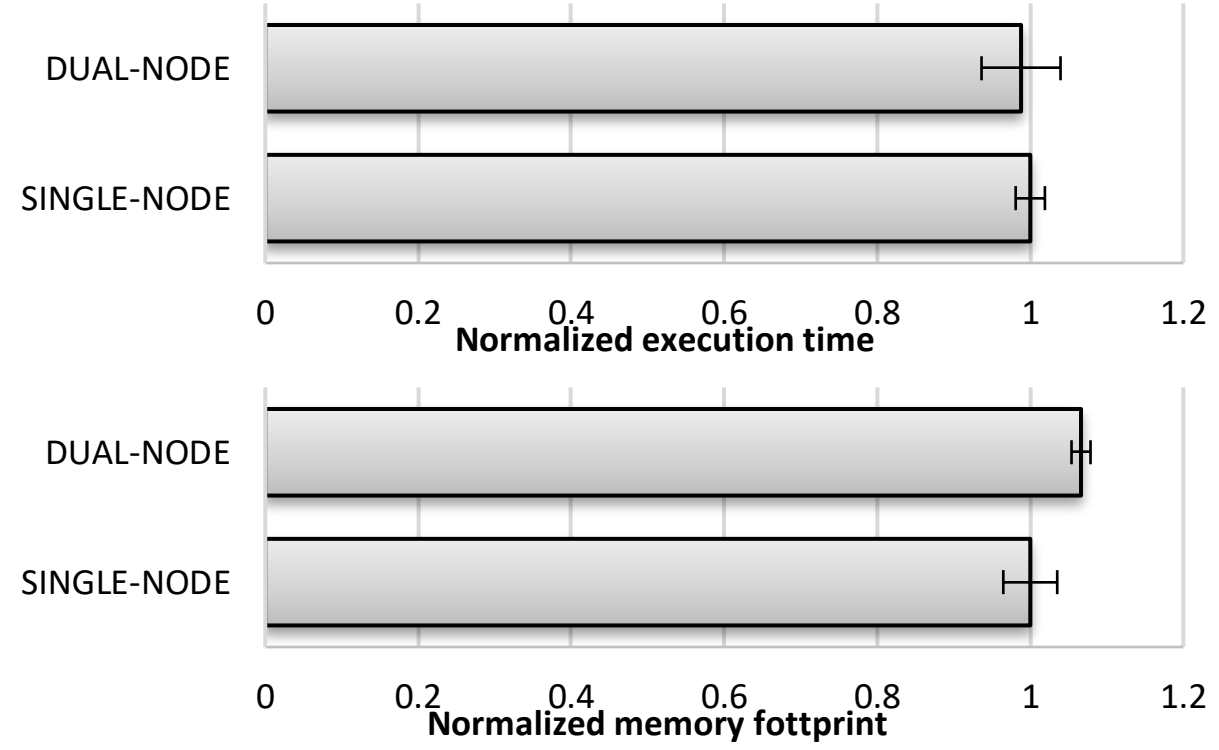

**Figure 12. Dual-node simulation overhead. Execution time (up) and memory footprint (down) of the simulation. Normalized results.**

A similar result is obtained from the analysis of the amount of memory consumed by each kind of simulation. This memory footprint joins gem5 memory and the amount of main memory assigned to simulated machines, both client and server. For our experiments, the amount of memory devoted to the client node was reduced to 1GB (the minimum required to run YCSB and SPECweb clients), in contrast to the 4GB of the server node. Figure 12.down shows that the additional 1GB required by client node only translates into less than 10% memory consumption overhead, where the most consuming components is the server node and the simulator itself.

# 8 CONCLUSIONS

In this work we have carried out a group of experiments that illustrate the relevance of choosing the appropriate simulation methodology. We showed that the error induced by the combined execution of distributed workloads on a single-node is far from negligible in some cases. We conclude that more suitable methodologies are required to work with workloads suitable for multi-node environments, providing Gem5 the required features for such a complex task (multi-node simulation support as well as hardware assisted simulation). Making use of these features we conduct a set of simulation experiments that confirm the unsuitability of single-node evaluations of distributed (client-server or multi-node) workloads.

Our next step to extend our conclusions will consist of adapting additional benchmark suites for gem5 simulations in order to analyze a broader range of environments. We also plan to extend our evaluation to more architectural components where interference could lead to incorrect results, such as L1 Instruction cache or branch predictor.

# 9 ACKNOWLEDGEMENTS

This work was supported in part by the Spanish Government (Secretaría de Estado de Inverstigación, Desarrollo e Innovación) under Grants TIN2015-66979-R and TIN2016-80512-R.